\documentstyle[preprint,aps]{revtex}
\begin{document}                % INITIALIZE - DONT CHANGE
\draft
\title{Measuring surface-area-to-volume ratios in soft porous materials
using laser-polarized  xenon interphase exchange NMR}
\author{J. P. Butler,$\dagger^{1,2}$ R. W. Mair,$^{3}$ S. Patz,$^{2}$ D.
Hoffmann,*$^{3}$ M. I. Hrovat,$^{4}$ R. A. Rogers,$^{1}$ G. P. Topulos$^{2}$
and R. L. Walsworth$^{3}$}
\address{$^{1}$Harvard School of Public Health, Boston, MA 02115, USA}
\address{$^{2}$Brigham and Women's Hospital and Harvard Medical School,
Boston, MA 02115, USA}
\address{$^{3}$Harvard-Smithsonian Center for Astrophysics, Cambridge,
MA 02138, USA}
\address{$^{4}$Mirtech Inc., Brockton, MA 02301, USA}
\address{$\dagger$ To whom correspondance should be addressed}
\address{* Present address:  Focused Research, Madison, WI 53711, USA}
\date{\today}
\maketitle
\begin{abstract}                % DON'T CHANGE THIS LINE
We demonstrate a minimally invasive nuclear magnetic resonance (NMR) technique
that enables determination of the surface-area-to-volume ratio ($S/V$) of soft
porous materials from measurements of the diffusive exchange of laser-polarized
$^{129}$Xe between gas in the pore space and $^{129}$Xe dissolved in the solid
phase.  We apply this NMR technique to porous polymer samples and find
approximate agreement with destructive stereological measurements of $S/V$
obtained with optical confocal microscopy.  Potential applications of
laser-polarized xenon interphase exchange NMR include measurements of in vivo
lung function in humans and characterization of gas chromatography columns.
\end{abstract}
%
% body of paper here
%
Porous media are ubiquitous in nature, e.g., granular materials, foams,
ceramics, oil- or water-bearing ``reservoir'' rocks, and animal and human
lungs.  Determining the structure of these materials is relevant to a wide
range of scientific and technological problems, ranging from the coarsening
of foams to the transport properties of subsurface fluids to cardiopulmonary
physiology and medicine.  In this paper we demonstrate a noninvasive
technique to characterize the the surface-area-to-volume ratio ($S/V$)
\cite{1} of ``soft'' porous media: i.e., materials in which there is
significant gas solubility in the solid phase.  Such a description applies to
many different materials, including porous polymer granulates used for
filtering, porous polymer bead packs used for radio-immunoassay, resin
columns used for chromatographic separation, and lung and sinus tissue in
humans and animals.

$S/V$ influences numerous interactions within porous media, including fluid
diffusion and transport, electric charge distribution, and chemical
exchange.  For example, $S/V$ is important in determining pulmonary function
since the lung is the site of O$_2$ and CO$_2$ exchange between the body and
the external environment.  There are numerous techniques to measure $S/V$ in
porous media, including stereology \cite{2} with traditional and confocal
microscopy, light scattering \cite{4,5}, and mercury intrusion porosimetry
\cite{5}.  However, only nuclear magnetic resonance (NMR) provides a
noninvasive and nondestructive technique for determining 3D structure,
including $S/V$ \cite{6}.

In particular, NMR measurements of the time-dependent diffusion coefficient
of a liquid filling the pore space has been used to determine $S/V$ in a
variety of porous media \cite{7,8} including glass beads \cite{9}, reservoir
rocks \cite{10}, and biological samples \cite{11}.  Recently, we extended this
technique to NMR measurements of xenon gas diffusion within porous media
\cite{12,13}, thereby probing longer length scales than with traditional
liquid-infused diffusion measurements \cite{9,10} because of fast gas
diffusion and long xenon spin polarization lifetimes.  Gas diffusion NMR may
also be used to study porous media that are not compatible with
water-saturation, such as non-wetting polymers and in vivo lungs.  However,
for materials with pore size less than about 300 $\mu\text{m}$ (such as lung
alveoli), the high diffusivity of gases causes systematic errors in the
determination of $S/V$ because the rms diffusion distance during the
application of RF and magnetic field gradient pulses is comparable to the
mean pore size \cite{12,13,14}.  Reducing this effect with faster gradient
pulses, buffer gases, or higher gas pressures is impractical \cite{15}.

Thus we have devised an alternative NMR method (``xenon interphase exchange
NMR'') to measure $S/V$ in soft (i.e., xenon-soluble) porous media.  In this
technique we determine $S/V$ from the rate of $^{129}$Xe polarization (not
chemical) exchange between gas in the pore space and $^{129}$Xe dissolved in
the solid.  First, we infuse the sample with laser-polarized (i.e., highly
spin-polarized or hyperpolarized) $^{129}$Xe gas.  Next, we rapidly destroy
(``quench'') all dissolved-phase $^{129}$Xe polarization using a suitable
sequence of RF and magnetic field gradient pulses.  The result is an initial
condition of NMR-labeled (i.e., polarized) $^{129}$Xe in the gas phase with
unpolarized $^{129}$Xe in the dissolved phase.  We then measure the diffusion
of $^{129}$Xe polarization into the dissolved phase by monitoring the recovery
of the NMR signal at the chemically-shifted frequency of dissolved-phase
$^{129}$Xe.  The rate of increase of the dissolved-phase $^{129}$Xe NMR
signal is proportional to $S/V$ (see analysis below).

Fast, high-resolution NMR measurements of the nuclear spin-$\frac{1}{2}$
noble gases $^{3}$He and $^{129}$Xe are possible with optical pumping
techniques using high-power lasers which increase the noble gas spin
polarization to $\sim 1-10\%$ \cite{16}.  Laser-polarized noble gas NMR is a
powerful technique in the physical and biomedical sciences, e.g., in lung
imaging \cite{17}.  For interphase exchange measurements, $^{129}$Xe is used
rather than $^{3}$He because of two factors.  First, $^{129}$Xe dissolves
about ten times more readily than $^{3}$He in lipophilic compounds: e.g.,
$^{129}$Xe has a large dissolved/gas partition coefficient of $\sim$ 0.25 for
lung tissue \cite{18}.  Second, the gas and dissolved-phase $^{129}$Xe NMR
frequencies are typically separated (chemically-shifted) by $\sim$ 200 ppm, a
relatively large value (the typical $^{3}$He chemical shift is less than 1
ppm) which makes it easy to manipulate or observe $^{129}$Xe spins
selectively in either phase \cite{19}.

Xenon interphase exchange NMR may be particularly important for the human
lung where the mean pore (alveolar) size is $\sim$ 100 $\mu\text{m}$
\cite{3,4}.  A noninvasive method to measure and image $S/V$ and gas exchange
in the human lung could be an important tool for probing lung physiology and
pathophysiology \cite{20}; e.g., in the diagnosis and treatment of asthma and
chronic obstructive pulmonary disease (COPD), the fourth leading cause of
death in the U.S \cite{21}.  Traditional lung function tests of gas exchange
parameters are seriously limited by the inability to provoke step changes in
gas composition or properties deep in the lung.  Deconvolving the effects of
ventilatory dilution, tissue absorption, and blood perfusion is a major
challenge.  By contrast, xenon interphase exchange NMR provides an almost
ideal step change in $^{129}$Xe polarization between tissue and gas, and thus
offers a unique window into lung structure and function.  For example,
subsequent to our annoucement of preliminary results using xenon
interphase exchange NMR \cite{22}, Ruppert et. al. succesfully applied a
similar technqiue to studies of lung function in dogs \cite{23}. Xenon
interphase exchange NMR may also be a powerful tool in the characterization of
non-living, soft porous media.  For example, knowledge and control of $S/V$ in
packed columns of polymeric beads - commonly used in molecular fractionation
applications - is critical to achieving adequate molecular separation.

In the study reported here, we monitored the exchange of $^{129}$Xe
polarization from the gas phase into the dissolved phase in a xenon-soluble
porous polymer granulate (POREX Corp., Atlanta, GA), chosen as a model (with
respect to mean pore size) for lung tissue.  We investigated four different
samples, which the manufacturer stated as having median pore sizes of 20, 70,
120, and 250 $\mu\text{m}$ (obtained with mercury intrusion porosimetry on
random production samples) and similar porosities of $\sim 30\%$.  The
interphase exchange experimental protocol was as follows.  First, we filled a
glass container completely with a porous polymer sample, excluding free gas
outside the cylindrical polymer core.  Next, we placed the sample in an NMR
magnet, pumped out the air from the sample's pore space, and allowed
laser-polarized xenon gas to fill the space.  (NMR imaging showed that the
xenon gas uniformly filled the pore space within a few seconds.)  We then
used a frequency-selective RF pulse to quench the dissolved-phase $^{129}$Xe
polarization, and observed the recovery of the dissolved-phase $^{129}$Xe NMR
signal due to interphase exchange for times from 10 to 1000 ms.  The NMR
pulse sequence for the interphase exchange measurement is shown schematically
in Fig.~\ref{Fig1}.  Note that we made an initial measure of the gas-phase
$^{129}$Xe NMR signal using an RF pulse with a small flip angle $\alpha  \sim
3 - 5\,^{\circ}$.  Also, the selective RF pulse at the dissolved-phase
$^{129}$Xe NMR frequency was not perfect and hence generated a small NMR
signal from $^{129}$Xe gas in the pores.  We exploited these gas-phase signals
to correct the data for slowly decreasing $^{129}$Xe polarization due to both
RF depletion and $T_1$ relaxation in the pores.  An example of the observed
interphase exchange spectra is shown in Fig.~\ref{Fig2}.  We performed xenon
interphase exchange experiments on two NMR spectrometers at different
magnetic field strengths: a horizontal-bore GE Omega/CSI spectrometer
operating at 4.7 T (55.3 MHz $^{129}$Xe NMR frequency) with field gradient
strength up to 7 G/cm; and a modified SMIS spectrometer interfaced to a
prototype IBM horizontal-bore magnet operating at 1.5 T (17.6 MHz $^{129}$Xe
NMR frequency) with field gradient strength up to 2 G/cm.  For each
experimental run, we used the spin-exchange optical pumping technique
\cite{16} in the fringe field of the NMR magnet and achieved $\sim 0.5\%$
$^{129}$Xe spin polarization for ~ 2 bar-liters of xenon gas (natural isotopic
abundance).

Calculation of absolute values for $S/V$ required prior characterization of the
samples.  We used thermally-polarized xenon NMR on both spectrometers to
determine the porosity $\phi$ and $^{129}$Xe partition coefficient $b$ for
the four porous polymer samples.  We placed all 4 sample cells on a manifold
along with an empty cell of equal volume, admitted $\sim$ 3 bar of xenon gas
(natural isotopic abundance) and $\sim$ 2 bar of oxygen, and allowed the
system to equilibrate over 2 - 3 days to achieve full penetration of xenon
into the polymer samples.  We then measured the $^{129}$Xe NMR gas signal in
the empty cell ($s_{g0}$) as well as the equilibrium gas and dissolved-phase
$^{129}$Xe NMR signals ($s_{gas}, s_{diss}$) in the four polymer samples.  For
each sample, $\phi = s_{gas}/s_{g0}$ and $b = (s_{diss}/s_{gas})\phi/(1 -
\phi)$. We obtained similar results for $\phi$ and $b$ at 4.7 T and 1.5 T.
Measurement of the dissolved-phase $^{129}$Xe diffusion coefficient $D_{diss}$
required large magnetic field gradients, which was only possible across small
sample dimensions.  Hence we cut additional pieces of each polymer sample to
fit in standard 5 mm diameter NMR spectroscopy tubes and filled the tubes with
$\sim$ 10 bar xenon gas (isotopically-enriched to $90\% ^{129}$Xe). We then
measured $D_{diss}$ using a 9.4 T Bruker AMX2 spectrometer with 50 G/cm field
gradients, employing a modified pulsed-gradient technique previously applied
to gas-phase diffusion measurements on rocks and glass beads \cite{12}. 
Table~\ref{Tab1} lists the measured values of $\phi, b$, and $D_{diss}$, along
with the $^{129}$Xe longitudinal spin relaxation times ($T_1$) for each porous
polymer sample.

To determine $S/V$ from the xenon interphase exchange NMR measurements, we
employed a one-dimensional diffusion model, asymptotically valid for short
times.  We assumed that the density of dissolved-phase $^{129}$Xe polarization,
$\rho_{diss}(x,t)$, satisfies the diffusion equation on a half space, $x\ >\,
0$, where the non-trivial spatial coordinate $x$ is normal to the gas/solid
interface. The quench of the dissolved-phase $^{129}$Xe polarization created
the initial condition $\rho_{diss}(x,0) = 0$, $x\ >\, 0$. Given the much
greater gas-phase $^{129}$Xe density, we also assumed a fixed boundary
condition $\rho_{diss}(0,t) = b\rho_{gas}$.  Here, $\rho_{gas}$ is the density
of gas-phase $^{129}$Xe polarization, which we found to be constant during an
individual interphase exchange measurement except for the small effects of RF
depletion and slow $T_1$ relaxation mentioned above.  The solution of the
diffusion equation for $\rho_{diss}(x,t)$ is given by the well known error
function \cite{24}.  The number of dissolved-phase polarized $^{129}$Xe atoms
$N_{diss}(t)$ can then be obtained by direct integration of the density
or by solving the ordinary differential equation obtained by integrating the
diffusion equation.  The result is:
\begin{equation}
\label{eq:1}
\frac{N_{diss}(t)}{V} = \rho_{gas}b\frac{S}{V}\sqrt{4D_{diss}t/\pi}
\end{equation}
where $S$ is the total pore-space surface area of the sample, and the
Euclidean volume $V$ of the sample has been introduced as a common divisor on
both sides of the equation.  Noting that
\begin{equation}
\label{eq:2}
\rho_{gas} = \frac{N_{gas}}{\phi V}
\end{equation}
where $N_{gas}$ is the number of gas-phase polarized $^{129}$Xe atoms (assumed
constant except for RF depletion and $T_1$ relaxation), Eq. (1) can be
rewritten as
\begin{equation}
\label{eq:3}
\frac{N_{diss}(t)}{N_{gas}} = \frac{bS}{\phi V}\sqrt{4D_{diss}t/\pi}.
\end{equation}
$N_{diss}(t)$ and $N_{gas}$ are proportional to the frequency integrals over
the observed $^{129}$Xe NMR spectral peaks, $s_{diss}(t,\omega)$ and
$s_{gas}(\omega)$, for the dissolved and gas phases respectively.  Thus we
have
\begin{equation}
\label{eq:4}
I(t) = \frac{\int s_{diss}(t,\omega)d\omega}{\int s_{gas}(\omega)d\omega} =
\frac{bS}{\phi V}\sqrt{4D_{diss}t/\pi}.
\end{equation}
We determined $S/V$ by fitting the ratio of dissolved and gas-phase
$^{129}$Xe NMR signals to $\sqrt{t}$ and using measured values for $b, \phi$
and $D_{diss}$

Concurrent with the NMR studies, we also characterized the four samples of
polymer granulate using confocal microscopy in order to give independent
measurements of $S/V$.  We carefully sliced sections of each of the four
polymer rods with different pore sizes, creating glass smooth surfaces of both
the longitudinal and cross-sectional faces of the rods.  We oriented these
sectional samples on a Leica TCS NT confocal microscopeconfigured for 488 nm
excitation and reflected light imaging \cite{25}.  The resulting microscope
images showed the polymer granulates as solids, and the pore spaces as black
voids.  Computer-based image analysis (ImageSpace, Molecular Dynamics,
Sunnyvale, CA) superimposed multiple, equally-spaced lines across the images,
and interface boundaries on each line of the images were counted.  Thus for
each sample we determined the mean linear intercept, $L_m$, which is defined as
the average distance  between phase boundaries in an image.  While $L_m$ is but
one of several possible measures of pore size, it has the unique and powerful
advantage of being stereologically related to the surface to volume ratio. In
particular, $L_m$ is an unbiased estimator of $S/V$ in the image plane: $L_m =
2(S/V)^{-1}$ \cite{26}.  See Table~\ref{Tab2} for the optically-determined
values of $L_m$ and $S/V$.

We acquired three xenon interphase exchange data sets for each of the four
different pore size polymer samples: two data sets at 4.7 T and one at 1.5
T.  As an example, Fig.~\ref{Fig3} shows an interphase exchange data set
taken on the 4.7 T instrument.  Each of the points plotted in Fig.~\ref{Fig3}
consists of the normalized average of three or four separate series of
exchange spectra of the type shown in Fig.~\ref{Fig2}.  The normalized xenon
integral displayed on the ordinate of Fig.~\ref{Fig3} is given by
$\langle I(t) \rangle \phi sin \alpha/b$ (see Eq. (4)), where $\alpha$ is the
small flip angle of the RF pulse used in the initial gas-phase interrogation. 
All xenon interphase exchange NMR data followed the expected $\sqrt{t}$ trend
over the observed range of exchange times (out to $\sim$ 1 s).

Fig.~\ref{Fig4} compares $S/V$ values derived from the optical and NMR
techniques for each of the four polymer samples: one optical and three NMR
data sets.  (See also Table~\ref{Tab2}.)  The solid line in Fig.~\ref{Fig4}
is the line of identity for the optical and NMR values of $S/V$; the dashed
line is a line of regression constrained to go through the origin.  The slope
of the regression line is 1.27 $\pm$ 0.18, indicating approximate agreement
between the optical and NMR measures of $S/V$.  Sources of systematic error
in the interphase exchange NMR measurement include: miscalibration of the
flip angle $\alpha$ of the gas-phase interrogation RF pulse; inaccuracy in the
dissolved-phase xenon diffusion coefficient measurements; and deviations from
the simple, one-dimensional diffusion model used to derive $S/V$.  For
example, the very low dissolved-phase xenon NMR signal from the nominally 120
$\mu$m POREX sample prevented us from measuring $D_{diss}$ for this sample.
Instead, we used the average of the measured $D_{diss}$ values for the
nominally 70 $\mu$m and 250 $\mu$m samples to calculate $S/V$ from the
interphase exchange measurements for the 120 $\mu$m sample.

We note that the manufacturer's quoted pore sizes for the four porous polymer
samples do not agree with our pooled optical and NMR measurements of $S/V$
(and hence $L_m$); even the rank order does not agree.  This discrepancy may
not be surprising, since the pore sizes given by the manufacturer were
determined by mercury intrusion porosimetry \cite{5,27}, which is primarily a
probe of minimum pore-throat size and thus not necessarily a good measure of
$L_m$ or $S/V$ \cite{26}.  We emphasize that there are different choices of
the definition of pore size; which choice is most appropriate will depend on
the particular application.

In conclusion, we have demonstrated a new technique, xenon interphase
exchange NMR, that provides a minimally invasive and nondestructive probe of
the surface-area-to-volume ratio ($S/V$) of xenon-soluble porous materials.
We have applied this technique to porous polymer samples and found approximate
agreement with destructive stereological measurements of $S/V$ made with
optical confocal microscopy.

We thank Prof. David Cory for access to the 9.4 T NMR instrument, and Dr.
Werner Maas for the loan of a gradient-equipped x-nucleus probe.  We also
acknowledge Ben Hirokawa at POREX Corp. who provided us with porous polymer
granulate samples.  This work was funded by NASA grant NAG9-1166, NIH grants
R21-RR/CA14297 and R01 HL-55569, and the Smithsonian Institution Scholarly
Studies Program.

%Table 1
\begin{table}
\caption{For each porous polymer sample, measured values are listed for the
porosity $\phi$, the $^{129}$Xe gas/polymer partition coefficient $b$, and the
dissolved-phase $^{129}$Xe diffusion coefficient $D_{diss}$.  Also listed are
the $^{129}$Xe longitudinal spin relaxation time (polarization lifetime) in
the dissolved phase ($T_{1diss}$) and the pore gas phase ($T_{1gas}$). The 
nominal pore size is that provided by the polymer manufacturer from mercury
intrusion porimetry measurements on random production samples. Note:
$T_{1diss}$ was the same at 4.7 T and 1.5 T; also, $D_{diss}$ was not
measured in the nominally 120 $\mu$m sample because the dissolved-phase
$^{129}$Xe NMR signal from this sample was too low to allow the experiment to
be completed in a reasonable time. Listed measurement uncertainties do not
include estimates of systematic errors.}
\label{Tab1}
\begin{tabular}{|l|r|r|r|r|}
\textbf{Nominal pore size ($\mu$m)} & \textbf{20} & \textbf{70} & \textbf{120}
& \textbf{250} \\
\hline \hline
Porosity, $\phi$ & 0.45 $\pm$ 0.02 & 0.45 $\pm$ 0.03 & 0.46 $\pm$ 0.01 & 0.43
$\pm$ 0.02 \\
\hline
Xe partition coefficient, $b$ & 0.64 $\pm$ 0.02 & 0.52 $\pm$ 0.04 & 0.56 $\pm$
0.03 & 0.63 $\pm$ 0.03 \\
\hline
Xe $D_{diss}$ ($\times 10^{-12}$ m$^2$s$^{-1}$) & 5.6 $\pm$ 0.5 & 7.0 $\pm$ 0.7
& n/a & 7.2 $\pm$ 0.7 \\
\hline
Xe $T_{1diss}$ (s) at 4.7 T  & 8.1 $\pm$ 0.4 & 7.7 $\pm$ 0.3 & 7.6 $\pm$ 0.3 &
8.0 $\pm$ 0.3 \\
\hline
Xe $T_{1gas}$ (s) at 1.5 T & 21.8 $\pm$ 0.5 & 34.1 $\pm$ 0.5 & 35.4 $\pm$ 0.4 &
19.9 $\pm$ 0.2 \\
\hline
Xe $T_{1gas}$ (s) at 4.7 T & 41.6 $\pm$ 0.6 & 68.1 $\pm$ 1.4 & 70.0 $\pm$ 1.6 &
42.8 $\pm$ 0.6 \\
\end{tabular}
\end{table}

%Table 2
\begin{table}
\caption{The surface-area-to-volume ratios ($S/V$) derived from the optical
and NMR measurements are listed for each of the four polymer samples.  Also
listed is the mean linear intercept, $L_m = 2(S/V)^{-1}$, determined from the
optical microscopy measurements. Listed measurement uncertainties do not
include estimates of systematic errors.}
\label{Tab2}
\begin{tabular}{|l|r|r|r|r|}
\textbf{Nominal pore size ($\mu$m}) & \textbf{20} & \textbf{70} & \textbf{120}
& \textbf{250} \\
\hline \hline
$S/V$ from xenon NMR ($\mu$m$^{-1}$) & 0.030 $\pm$ 0.006 & 0.016 $\pm$ 0.005 &
0.016 $\pm$ 0.003 & 0.018 $\pm$ 0.003 \\
\hline
$S/V$ from microscopy ($\mu$m$^{-1}$) & 0.0229 $\pm$ 0.0004 & 0.0135 $\pm$
0.0003 & 0.0122 $\pm$ 0.0003 & 0.0127 $\pm$ 0.0003 \\
\hline
$L_m$ from microscopy ($\mu$m) & 87.3 $\pm$ 1.7 & 147.8 $\pm$ 3.9 & 164.3
$\pm$ 4.5 & 158.0 $\pm$ 4.3 \\ 
\end{tabular}
\end{table}

%Figure 1
\begin{figure}
\caption{Xenon interphase exchange NMR pulse sequence.  The sequence can be
divided into the following segments.  A: interrogation of the gas-phase NMR
signal ($s_{gas}$) with a small flip angle RF pulse; $s_{gas}$ is
proportional to the $^{129}$Xe polarization.  B: semi-selective saturation
train of RF pulses to destroy the dissolved-phase $^{129}$Xe polarization.  C:
interphase exchange time, $t$.  D: semi-selective $90\,^{\circ}$ RF pulse to
observe the recovery of the dissolved-phase $^{129}$Xe signal ($s_{diss}(t)$);
this RF pulse also excites a small gas-phase signal, allowing correction for
incremental loss of gas-phase magnetization.  E: interphase exchange
measurement, typically repeated ten times for each exchange time, $t$.}
\label{Fig1}
\end{figure}

%Figure 2
\begin{figure}
\caption{Dissolved-phase xenon NMR spectra as a function of interphase
exchange time for the nominally 20 $\mu$m pore-size sample.  The peaks at 0
ppm arise from the weak, far-off-resonance excitation of gas-phase polarized
$^{129}$Xe when dissolved-phase $^{129}$Xe is pulsed semi-selectively.
Following saturation, the dissolved-phase $^{129}$Xe peaks at 200 ppm show
recovery as a function of exchange time.  The observed spectra are scaled to
normalize the integrals of each gas peak to the gas peak in the first
exchange spectrum; this normalization accounts for the slow loss of gas-phase
magnetization due to exchange, RF depletion, and $T_1$ relaxation.}
\label{Fig2}
\end{figure}

%Figure 3
\begin{figure}
\caption{One set of xenon interphase exchange NMR data, acquired at 4.7 T for
each of the four porous polymer samples.  (The samples are indicated in the
figure by the manufacturer's claimed pore size.)  Each data point represents
the weighted mean of 3 to 4 normalized spectra of the type illustrated in
Fig.~\ref{Fig2}, acquired in repeated runs of the experiment.  The normalized
xenon integral plotted on the ordinate is thus given by $\langle I(t) \rangle
\phi sin \alpha/b$ (see Eq. (4)).  Uncertainties for each interphase exchange
spectrum were determined from the measured rms noise within the spectral
bandwidth, normalized for variations in gas-phase magnetization;
one-standard-deviation error bars were then calculated for the weighted means
of the normalized xenon integrals.  The error bars are larger for longer
exchange times because these data were acquired later in each experimental run
and hence had lower gas-phase magnetization, which correspondingly increased
the normalized noise of the dissolved-phase xenon spectra.}
\label{Fig3}
\end{figure}

%Figure 4
\begin{figure}
\caption{Comparison of polymer pore-space $S/V$ values obtained with optical
microscopy and NMR.  For each of the four polymer samples, three NMR
experimental runs are plotted versus one optical measurement.  The solid line
is a line of unity for the $S/V$ values provided by the two measurement
techniques.  The dashed line indicates a regression between the NMR and
optical $S/V$ values, constrained to pass through the origin; this regression
line has a slope of 1.27 $\pm$ 0.18.}
\label{Fig4}
\end{figure}

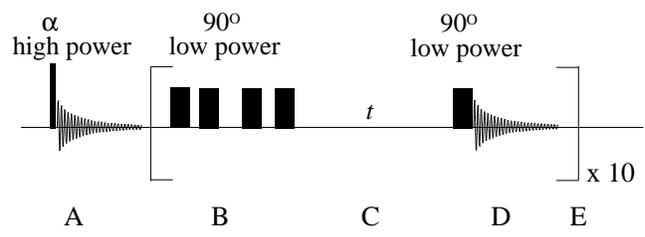
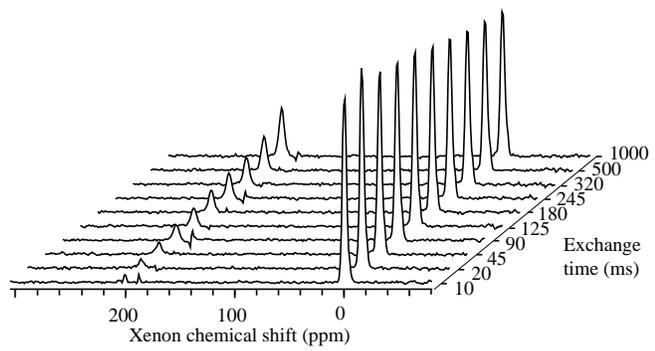
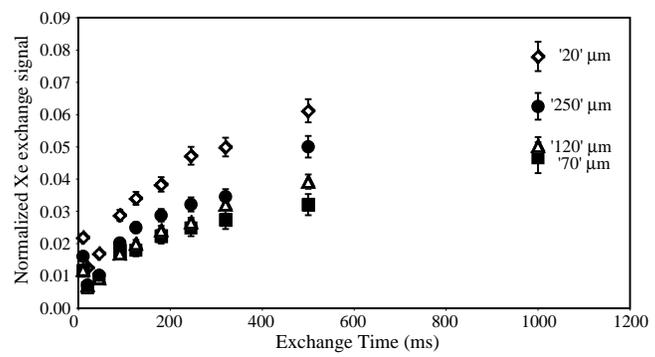
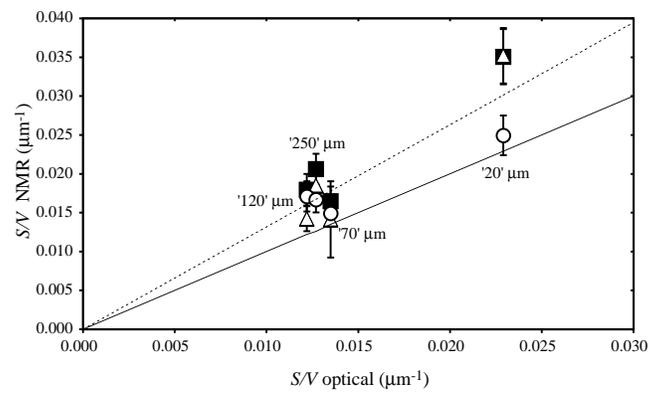
\end{document}